\documentclass[a4paper,fleqn,usenatbib]{mnras}
\pdfoutput=1
\usepackage{newtxtext,newtxmath}

\usepackage[T1]{fontenc}
\usepackage{ae,aecompl}

\usepackage{graphicx}	
\usepackage{amsmath}	
\usepackage{amssymb}	
\usepackage{gensymb}

\title[High-contrast imaging of HD 163296]{High-contrast imaging of HD 163296 with the Keck/NIRC2 L$^\prime$-band vortex coronograph}
\author[G. Guidi et al.]{
G. Guidi,$^{1,2,3}$\thanks{E-mail: guidi@arcetri.inaf.it}
G. Ruane,$^{4,5}$
J. P. Williams,$^{3}$
D. Mawet,$^{4,6}$
L. Testi,$^{1,7}$
\newauthor{
A. Zurlo,$^{8,9}$
O. Absil,$^{10,11}$
M. Bottom,$^{6}$
\'{E}. Choquet,$^{4,6,12}$
V. Christiaens,$^{10,13}$
}
\newauthor{
B. Femen\'ia Castell\'a,$^{14}$
E. Huby,$^{10,15}$
A. Isella,$^{16}$
J. Kastner,$^{17,18}$
T. Meshkat,$^{4,19}$
}
\newauthor{
M. Reggiani,$^{10}$
A. Riggs,$^{4,6}$
E. Serabyn,$^{4,6}$
N. Wallack$^{20}$}
\\
$^{1}$INAF-Osservatorio Astrofisico di Arcetri, Largo E. Fermi 5, I-50125 Firenze, Italy \\
$^{2}$Dipartimento di Fisica e Astronomia, Universit\`a degli Studi di Firenze, Italy \\
$^{3}$Institute for Astronomy, University of Hawai\'i at Manoa, Honolulu, HI 96822, USA \\
$^{4}$Department of Astronomy, California Institute of Technology, Pasadena, CA 91125, USA\\
$^{5}$NSF Astronomy and Astrophysics Postdoctoral Fellow\\
$^{6}$Jet Propulsion Laboratory, California Institute of Technology, Pasadena, CA 91109, USA\\
$^{7}$ESO, Karl Schwarzschild str. 2, D-85748 Garching bei Muenchen, Germany\\
$^{8}$N\'ucleo de Astronom\'ia, Facultad de Ingenier\'ia, Universidad Diego Portales, Av. Ejercito 441, Santiago, Chile\\
$^{9}$Escuela de Ingenier\'ia Industrial, Facultad de Ingenier\'ia y Ciencias, Universidad Diego Portales, Av. Ejercito 441, Santiago, Chile \\
$^{10}$Space Sciences, Technologies and Astrophysics Research (STAR) Institute, Universit\'e de Li\`ege, Li\`ege, Belgium\\
$^{11}$ F.R.S.-FNRS Research Associate\\
$^{12}$ Hubble Fellow\\
$^{13}$Departamento de Astronom\'ia, Universidad de Chile, Casilla 36-D, Santiago, Chile\\
$^{14}$W. M. Keck Observatory, Kamuela, HI 96743, USA\\
$^{15}$LESIA, Observatoire de Paris, Universit\'e PSL, CNRS, Sorbonne Universit\'e, Univ. Paris Diderot, Sorbonne Paris Cit\'e, 5 place\\ Jules Janssen, 92195 Meudon, France\\
$^{16}$Department of Physics and Astronomy, Rice University, 6100 Main St., 77005, MS-108, Houston, Texas\\
$^{17}$ School of Physics \& Astronomy, Laboratory for Multi-wavelength Astrophysics, Rochester Institute of Technology, Rochester, NY 14623, USA\\
$^{18}$ Chester F. Carlson Center for Imaging Science, Rochester Institute of Technology, Rochester, NY 14623, USA\\
$^{19}$ IPAC, California Institute of Technology, Pasadena, CA 91125, USA\\
$^{20}$ Division of Geological and Planetary Sciences, California Institute of Technology, Pasadena, CA 91125, USA
}

\date{Accepted XXX. Received YYY; in original form ZZZ}

\pubyear{2017}

\begin{document}
\label{firstpage}
\pagerange{\pageref{firstpage}--\pageref{lastpage}}
\maketitle

\begin{abstract} 
We present observations of the nearby (D$\sim$100\,pc) Herbig star HD~163296 taken with the vortex coronograph at Keck/NIRC2 in the L' band (3.7~$\mu$m), to search for planetary mass companions in the ringed disc surrounding this pre-main sequence star. 
The images reveal an arc-like region of scattered light from the disc surface layers that is likely associated with the first bright ring detected with ALMA in the $\lambda$=1.3mm dust continuum at $\sim$65~au. We also detect a point-like source at $\sim$0\farcs5 projected  separation in the North-East direction, close to the inner edge of the second gap in the millimetre images. 
Comparing the point source photometry with the atmospheric emission models of non-accreting giant planets, we obtain a mass of 6--7~M$_J$ for a putative protoplanet, assuming a system age of 5~Myr. 
Based on the contrast at a 95\% level of completeness calculated on the emission-free regions of our images, we set upper limits for the masses of giant planets of 8--15~M$_J$, 4.5--6.5~M$_J$ and 2.5-4.0~M$_J$  at the locations of the first, second and third gap in the millimetre dust continuum, respectively. 
Further deep, high resolution thermal IR imaging of the HD~163296 system are warranted, to confirm the presence and nature of the point source and to better understand the structure of the dust disc.   
\end{abstract}




\section{Introduction}
Planetary systems form inside discs of gas and dust around pre-main sequence stars, within typical disc lifetimes of a few million years \citep[e.g.][]{2007ApJ...662.1067H,2010A&A...510A..72F}. Thanks to advances in the exoplanets detection techniques, $\sim$3700 exoplanets have been confirmed to date\footnote{source: Nasa Exoplanets Archive, http://exoplanetarchive.ipac.caltech.edu}. However, only a few have been directly imaged, as differentiating between the planet and the residual light from the parent star scattered by the atmosphere, and optical aberrations inside the instrument, even after adaptive optics correction 
represents an arduous technical challenge. 
Yet, direct imaging of exoplanets is a powerful technique that allows full spectroscopic, photometric and astrometric characterization of the planetary companions, and provides access to a wider range of planet-star separations ($\geq$ 5~au) compared to other methods \citep[see the review by][]{Fischer2014}. 
Since newly formed planets have a significant potential gravitational energy available, they are thought to have high temperatures and thus strong emission in the infrared range, which make them more easily detectable at these wavelengths. 
The number of imaged exoplanets has been growing in the last few years, with several companions found in young systems \citep[e.g.][]{Lagrange2010,Marois2010,Rameau2013,2015Sci...350...64M,2017A&A...605L...9C}. 

The interplay between planets and the disc in which they formed plays a major role in shaping the architecture of planetary systems as well as in determining the planets physical properties and orbital elements \citep{2014prpl.conf..667B}. These processes have been investigated theoretically through hydrodynamical simulations of disc-planet interaction \citep[see the review by][and reference therein]{2017arXiv170707148K}, 
but remain largely unexplored by observations. 
Therefore, detecting planets that are still embedded in a disc would provide new and important constraints on planet formation mechanisms, timescales, and locations in the disc. 

The recent development of new infrared detectors, high-contrast techniques and observing strategies combined with advanced adaptive optics systems
mounted on 8-10 meter class telescopes (e.g. VLT/NaCo, Gemini/GPI, VLT/SPHERE, Keck/NIRC2, LBT/LUCI) led to the first potential detections of planet candidates in nearby transitional discs; i.e. systems with inner cavities in the dust/gas distributions \citep{2014prpl.conf..497E}. 
However, the extensive post-processing required to optimize  high contrast observations 
is liable to generate false positives, and anisotropic scattering from dust in discs can be misinterpreted as planetary emission. 
For example, two candidate companions have been found around the transitional disc LkCa~15 \citep{2012ApJ...745....5K,2015Natur.527..342S}; however a inner disc component overlapping with the location of the protoplanet candidates was later detected in polarized light at the same location \citep{2015ApJ...808L..41T}. 
A candidate has been proposed for HD~169142 \citep{2014ApJ...792L..23R,2014ApJ...792L..22B}, but a successive detection in polarized light suggested the emission was due to scattered light from the disc atmosphere \citep{2018MNRAS.473.1774L};  
a previous detection in HD~100546 has been recently questioned by follow up observations by \citet{2017AJ....153..244R}, using GPI at Gemini South. 
Searches for planetary companions in non-transitional discs that present gaps or other structures in the dust emission have been attempted  \citep[e.g. HL~Tau and TWHya;][]{2015ApJ...812L..38T,2017AJ....154...73R}, but with no detections so far. 
However, the high sensitivity and contrast achieved by such innovative instruments provide the opportunity to put stringent upper limits on the mass of potential planetary candidates in protoplanetary discs.

\begin{table}
	\centering
	\caption{Stellar parameters for primary target HD~163296 and reference star HD~183665. $R$ mag: UCAC4 catalogue \citep{Zacharias2013}. W1 mag (3.4~$\mu$m): WISE catalogue \citep{Wright2010}.
 }
	\begin{tabular}{ccccc} 
		\hline
		Name & RA & DEC & $R_\textrm{mag}$ & W1\\
		\hline
		HD 163296 & 17 56 21.3 & -21 57 21.9 & 6.9 & 3.7\\
        HD 183665 & 19 31 24.6 & -21 02 24.3 & 6.4 & 3.5\\
		\hline
	\end{tabular}
    \label{tab:par}

	\centering
	\caption{Keck/NIRC2 Observations.}
	\begin{tabular}{ccccc} 
		\hline
     	UT Date & Target & $^a$T$_{int}$(s) & Exp. Time & PA rot.\\
        \hline
		2017 May 30 & HD 163296 & 0.18 & 41.3 min & 47.7$^\circ$ \\
        2017 May 31 & HD 163296 & 0.18 & 40.0 min & 40.4$^\circ$\\ 
		 & HD 183665 & 0.18 & 30.0 min & 30.7$^\circ$ \\
		\hline
	\end{tabular}
    \begin{flushleft}
    $^a$ Integration time per co-add, each frame was composed by 100 co-adds on the first night of observations (May 30), and by 150 co-adds on the second night (May 31).
    \end{flushleft}
    \label{tab:obs}
\end{table}

The star/disc system HD~163296 (MWC\,275) represents another excellent target for direct imaging planet searches. Its high brightness and large angular extent
makes it one of the best studied protoplanetary discs in the solar neighborhood.
HD\,163296 is a Herbig Ae star of spectral type A1 \citep{2001A&A...378..116M}, with stellar mass of $M_* = 2.3 M_{\odot}$, Luminosity of $L_* = 36 L_{\odot}$, and effective temperature of $T_\mathrm{eff}=9500K$, as computed by \citet{natta04}. 
Recently, a distance of 101.2$\pm$1.2\,pc was inferred from high precision measurements of the stellar parallax provided by the ESA GAIA mission \citep{2018arXiv180410121B}.  
Comparison with pre-main sequence evolutionary models indicates an age of 5$^{+0.3}_{-0.6}$~Myr \citep{2009A&A...495..901M}. 
The star is surrounded by a gas and dust-rich disc that has been detected in scattered light at large scales in optical \citep{2000ApJ...544..895G,2008ApJ...682..548W} and NIR wavelengths \citep{2010A&A...511A..74B,2008ApJ...678.1070S}. 
The disc extends to a radius of about 250~au in dust emission at $\sim$1~mm \citep{itziar}, while orbiting gas has been detected and mapped in a number of molecular species, including CO, H$_2$CO, HCN, HCO$^+$, N$_2$H$^+$ and various isotopologues \citep[e.g.][and references therein]{2015ApJ...813..128Q,2017A&A...606A.125S,2017A&A...605A..21C}. 
Recently, ALMA observations at high angular resolution ($\sim$0.2$^{\prime\prime}$) revealed a morphology of the continuum emission at 1.3 mm characterized by three dark rings with angular radii of 0.5, 0.8, and 1.3 arcsec \citep{Isella2016}, corresponding to orbital radii of about 50, 80, and 130~au, when accounting for the new distance measured by GAIA.  
A following work performing 2D hydrodynamical simulations of disc-planet interaction, coupled with 3D radiative transfer \citep{2018ApJ...857...87L}, found that the three observed gaps are consistent with half-Jovian mass planets (0.46~M$_J$, 0.46~M$_J$ and 0.58~M$_J$) partially depleting dust and gas around their orbits. 
Similarly, hydrodynamical simulations performed by \citet{2018arXiv180510290T} showed that two Jupiter-mass planets embedded in this disk in the second and third gap (83\,au and 137\,au) are consistent with the kinematical perturbations observed in the C$^{18}$O rotation pattern. 

In this paper we present the results of the observations of HD~163296 at L$^{\prime}$ band with the Keck/NIRC2 vortex coronograph. This observing program was aimed at directly detecting any young, massive planets that may be responsible for generating the rings/gaps features seen in the ALMA continuum images. 
We detect the HD~163296 disc in total intensity for the first time in the L$^{\prime}$ band: we see the scattered light from the inner ring of dust and identify a point-like source near the second dust gap we see in the ALMA image. 
Finally, the sensitivity of our observations allows us to set upper limits on the mass and accretion rates of giant planets present in this disc. 

\begin{figure}
\centering
\includegraphics[width=0.49\textwidth]{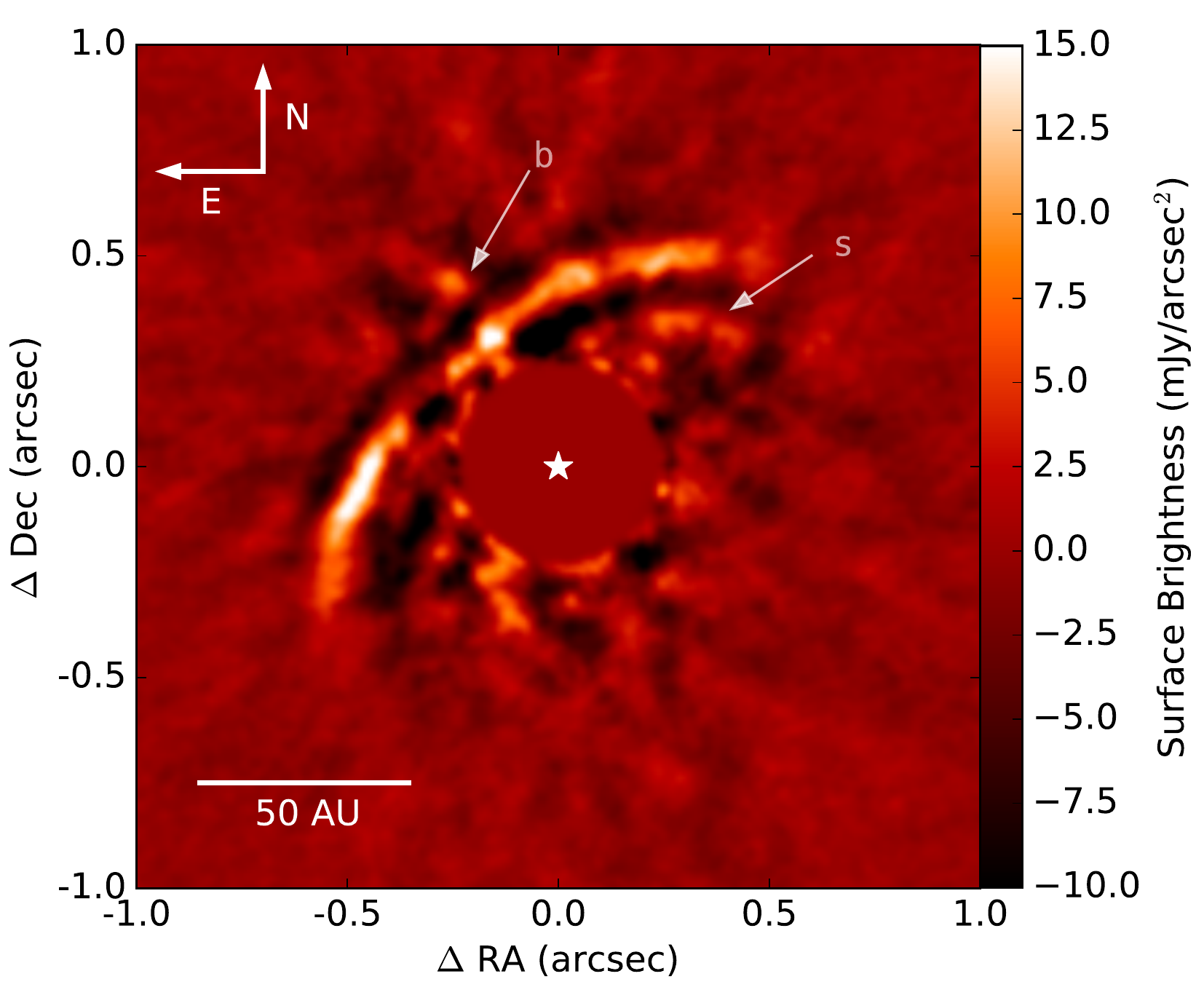}
\caption{Image of HD~163296 after two nights observations. The stellar contribution has been subtracted using PCA-ADI with 18 principal components computed from 204 total frames. A inner region with radius of 0\farcs24 (3*FWHM) was masked. The star symbol marks the position of HD~163296 at the centre of the frame.}
\label{fig:pca}
\end{figure}

\section{Observations and data reduction}
We observed HD\,163296 with Keck/NIRC2 vortex coronograph \citep{2017AJ....153...43S,2017AJ....153...44M} on two nights on UT 2017 May 30/31 at L$'$ band (central wavelength of 3.7$\mu$m) for $\sim$40 minutes each night. The pixel scale was 0.01 arcsec/pixel, and the angular resolution $\sim$0.08$\arcsec$. 
In order to apply the angular differential imaging (ADI) technique \citep{Marois2006} we tracked our target in pupil-stabilized mode, i.e. with the NIRC2 beam derotator fixed relative to the telescope elevation axis. 
On May 31, the reference star HD~183665 was also observed directly after the primary target to enable reference star differential imaging (RDI). The parameters of the two target stars are listed in Table \ref{tab:par}, while the integration times and parallactic angle rotations are reported in Table \ref{tab:obs}. 
Darks and sky flats were acquired during the same nights of observations, the average seeing was $\sim$0.6$\arcsec$ for both nights. The raw images were first reduced  
by subtracting the sky background and correcting for flat field and bad pixels, then co-registered and centered using the median speckle pattern. 

The science frames were acquired using the QACITS control loop \citep{2017A&A...600A..46H}, an automated observing sequence that allows pointing stability and a fine centering of the star on the vortex coronograph. 
The data were processed using the Vortex Image Processing (VIP) package \citep{2017AJ....154....7G}, applying Principal Component Analysis (PCA) with ADI and RDI. 
\section{Results}
\subsection{Total intensity image of HD~163296 after ADI processing}
\label{sec:res}
Figure \ref{fig:pca} shows the result of PCA-ADI PSF subtraction using 204 frames from both observing nights. We projected each frame onto 18 principal components (PCs) to build a model of the stellar speckle pattern over an annular region 0\farcs24-1.5$^{\prime\prime}$.
We see a clear detection of an elongated structure towards the North-East at a minimum distance of $\sim$0\farcs35 from the centre. The negative lobes appearing at each side are due to 
self-subtraction artefacts, typical of ADI-processed images. In particular, extended objects such as dust rings in discs can show alteration of several observable properties such as position angle, radial width, and total flux \citep{2012A&A...545A.111M}. 
Despite these possible distortion effects, we note that the geometry of the emission is overall consistent with the inclination and position angle of the disc as imaged in the sub-millimetre and millimetre continuum, corresponding to 42\degree and 132\degree, respectively, as found in \citet{Isella2016} and in agreement with previous studies \citep{2007A&A...469..213I,itziar,2015ApJ...813..128Q}.

An additional detection, spanning 0\farcs3-0\farcs5 in the north-west direction (labeled `s' in Fig.~\ref{fig:pca}), seems to follow a similar curvature, but the low signal-to-noise ratio (SNR) of this feature make it difficult to claim a reliable detection. 

Finally, a point-like source (labeled `b') appears in the North-East direction (PA=$30.6\degree\pm0.7\degree$) at $0\farcs49\pm0\farcs01$ from the central star with $\Delta L^\prime=11.1_{-0.2}^{+0.3}$, corresponding to an SNR of 4.7. The error bars and SNR were calculated by injecting and retrieving fake companions with the same flux and angular separation over a parallactic angle (PA) range 145\degree-305\degree (south-west region). This PA range was chosen to avoid confusing scattered light from the disc and speckle noise. 
As an additional test to confirm the physical nature of the `b'  point-like source, we show in Appendix \ref{app:b} its detection in a set of post-processed frames obtained running PCA-ADI with different parameters. 
We detected a background source at a larger separation that we classified as a background object, as described in Appendix \ref{app:a}. 

\subsection{Contrast limits}
We used the \texttt{contrast\_curve} method in the VIP package to determine the detection limits of our observations using a similar fake companion injection and retrieval method to that described in the previous section \citep{2017AJ....154....7G}.
The noise in the image was computed inside partial rings of $\lambda$/D resolution elements over a parallactic angle range 140\degree-310\degree. 
We computed an ensemble of detection limits using PCA-ADI with a varying number of PCs for three overlapping annular regions of the image: 0\farcs2-0\farcs8, 0\farcs3-1\farcs5, and 0\farcs5-2\farcs9 (2-9, 3-19, 5-37 $\lambda/D$, respectively). We also applied PCA-RDI in innermost region 0\farcs1-0\farcs5 (1-6 $\lambda/D$), but given the relatively small number of reference frames (<50\% of the science frames), RDI marginally improved the S/N and in a limited region at small separations (<0\farcs2). The corresponding RDI images were dominated by speckle- and PSF-residuals, therefore have not been used in the analysis. 
We used the combination of the overlapping annular regions described above to compute our contrast limits, as  performing PCA in annular subsamples of the frames helps the modelling and subtraction of the PSF residuals. 
Since no additional detection was obtained from the PCA reduction in the three separate regions, we only display in Fig.~\ref{fig:pca} the PCA-ADI frame obtained from the combination of outer radius and inner masking area that produces the best PSF subtraction in the area of interest for our analysis.   

\begin{figure}
\centering
\includegraphics[width=0.51\textwidth]{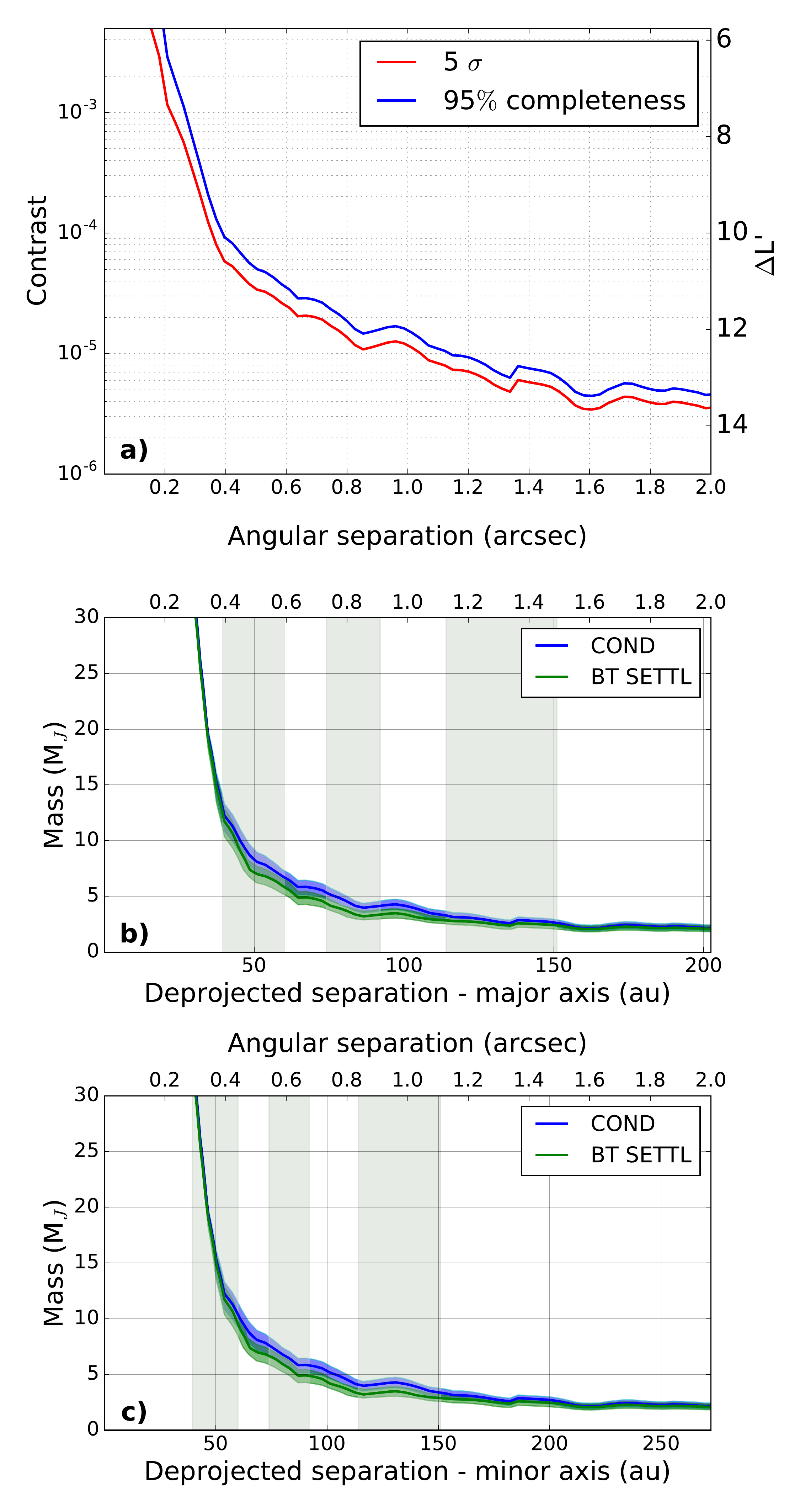}
\caption{\textit{Panel a)}: Contrast limit (planet-star flux ratio) after two observations of HD~163296, as a function of angular separation. Two cases of a 5 $\sigma$ level (red line) and a 95\% completeness (blue line) are shown. The curves are the combination of the contrast limits obtained applying PCA-RDI in the innermost regions (0\farcs1-0\farcs5, with 37 principal components), and PCA-ADI in three overlapping annular regions (0\farcs2-0\farcs8, 0\farcs3-1\farcs5 and 0\farcs5-2\farcs9 using 18, 15 and 36 PCs, respectively). 
\textit{Panel b) and c)}: planetary masses corresponding to a 95\% level of completeness, computed from the COND \citep{2003A&A...402..701B} and BT-Settl \citep{Allard2012} models assuming an age of 5$\pm$1~Myr. The shaded grey areas are drawn at the location of the gaps in the dust surface density as derived from modeling of $\lambda=$1.3~mm observations \citep{Isella2016}, at the deprojected separations along the disk major axis (panel b) and minor axis (panel c).}
\label{fig:cc}
\end{figure}

Figure \ref{fig:cc}a shows the resulting detection limits as a function of angular separation using a conventional 5$\sigma$ ``contrast curve" and an alternate method proposed by \citet{2018AJ....155...19J}. The 5$\sigma$ contrast curve assumes a Gaussian noise distribution and provides a flux ratio with a true positive fraction (TPF) of 0.5 and a fixed false positive fraction (FPF) of $2.9\times10^{-7}$. The alternate detection limits assume a Student-$t$ distribution, set a varying threshold that allows a 1\% chance of a false positive at evenly distributed within 1$^{\prime\prime}$, and provides TPF=0.95; i.e. 95\% completeness \citep[see e.g.][]{2018AJ....155...19J,2017AJ....154...73R}. The latter approach accounts for the small number of samples used to estimate the noise in the image, especially at small angular separations \citep{Mawet2014}. 

\begin{figure*}
\centering
\includegraphics[width=\textwidth]{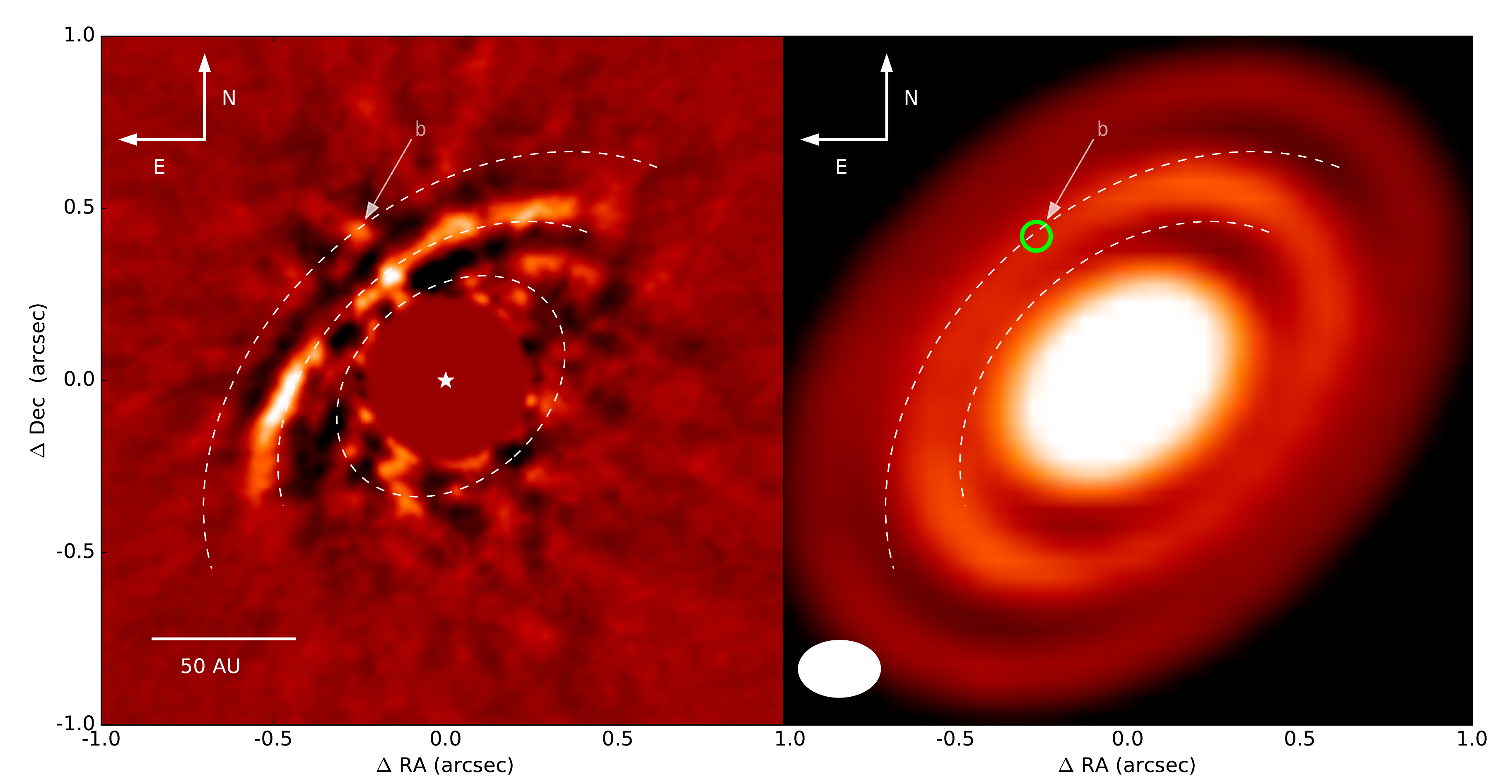}
\caption{\textit{Left panel}: Annotated NIRC2 image, also shown in Fig. \ref{fig:pca}. The dashed arcs of ellipse indicate scattering surfaces at different scale heights: the innermost ellipse is drawn at a scale height of 3~au above a location of 36~au on the midplane, the inner edge of the first gap in the dust distribution at $\lambda$=1.3~mm. 
The arc-like feature is likely generated from a layer 16 au above the inner wall of the first bright ring, at a location of 64 au on the midplane (intermediate arc of ellipse). A surface layer at a scale height of 28 au above the midplane distance of 94~au would be necessary to relate the point-like source to the inner edge of the second bright dust ring (outermost ellipse). \textit{Right panel}: the same arcs of ellipses are overlaid on the ALMA image showing the dust continuum at $\lambda=$1.3mm \citep{Isella2016} in logarithmic scale, with the beam ellipse drawn at the bottom left of the frame. If in the mid-plane, the position of the point source (labeled `b' and highlighted with a green circle) would be at a deprojected distance of 67~au, close to the inner edge of the second gap in the dust distribution.}
\label{fig:almakeck}
\end{figure*}

We translate the flux ratio at 95\% completeness into upper limits on planetary masses from luminosities computed for non-accreting giant planets by \citet{2003A&A...402..701B} and \citet{Allard2012}, using the stellar absolute flux  computed by WISE in the W1 filter (3.4\,$\mu$m), reported in Table~\ref{tab:par}. The resulting masses are shown in Fig. \ref{fig:cc}b and \ref{fig:cc}c.
Comparing the computed upper limits found for the planetary masses with the location of the gaps at mm wavelength showed in \citet{Isella2016}, in the inner most gap ($\sim$50~au) we would be sensitive to masses of about  8--15~M$_{J}$ along the direction of the major and minor axis, respectively. With the same criteria, in the second gap ($\sim$80~au) we find limits of 4.5--6.5~M$_{J}$ and in the third gap ($\sim$130~au) of 2.5--4.0~M$_{J}$. These planetary mass limits are much higher than the $\sim$0.5-1~M$_{J}$ planet expected to be clearing gaps in the dust and gas distribution based on the ALMA observations at $\lambda$=1.3~mm and hydrodynamical simulations \citep{2018ApJ...857...87L,2018arXiv180510290T}. However, the flux of any potential point source may also be explained by emission from an accreting circumplanetary disc \citep{Zhu2015}. Assuming such a putative circumplanetary disc has an inner radius of $\sim 1 R_J$, we constrain the product of the planet mass and mass accretion rate $M\dot{M}$ to $8.5\times10^{-7}$, $3.9\times10^{-7}$, and $2.1\times10^{-7}$~M$_J^2$/yr in the 50, 80, and 130~au gaps, respectively. 
Finally, we note that the derived upper limits for mass and accretion do not take into account the potential extinction by disc material, which can be significant in the inner regions of the system, and would result in an attenuation of the emission flux from a planetary companion.

Recently, the presence of a 2\,M$_J$ planet at a projected distance of 2\farcs3 and PA\,=\,-3$\degree$ has been proposed by \citet{2018arXiv180510293P} to explain a localized variation in the Keplerian velocity pattern of $^{12}$CO in HD~163296 as observed by ALMA. We do not detect any emission at this location, but since our estimated mass upper limits (shown in Figure \ref{fig:cc} up to a distance of 2\farcs0) at this position correspond to $\sim$2.2~M$_J$, 
we cannot exclude that a planetary body of about 2 Jupiter masses could remain undetected by our Keck observations.

\section{Discussion}

\subsection{Extended feature}
\label{sec:ext}
We interpret the bright arc-like feature on the North-East side of Fig. \ref{fig:pca} as scattered light from the disc atmosphere at the location of the first bright ring detected at mm wavelengths with ALMA by \citet{Isella2016}. 
In Fig. \ref{fig:almakeck} we present a side-by-side comparison of the Keck L' coronographic image, sensitive to  the dust reflecting the stellar radiation in the surface layers of the disc, 
and the ALMA continuum image at 1.3\,mm, which probes the dust thermal emission from the inner and colder regions of the disc midplane. 
The signal in the Keck image is detected only from the near side of the disc, found to be in the North-East direction from previous studies \citep[e.g.][]{2013ApJ...774...16R}. This points to an efficient forward scattering mechanism, and thus to large grains (with radius $a \gtrsim \lambda/2 \pi$) in the disc atmosphere \citep[see e.g.][]{Ardila2007,2011ApJ...738...23Q}. We note that a bright spot appears in the arc-like feature at the same position angle as the point-like source (PA=30.6\,\degree) labeled `b' in Fig.~\ref{fig:pca}. Both these features could be an effect of the scattering phase function being forward-peaked, thus  
producing a brighter signal close to the disc minor axis (PA=42\degree), on the near-side of the disc. 

The morphology of the structure in our ADI processed image is consistent with the $J$-band Gemini Planet Imager (GPI) image in polarized light \citep{2017ApJ...838...20M}. Based on a visual fit, \citet{2017ApJ...838...20M} determined the scattered light originates from a scale height of 18~au above the midplane at the location of 77~au (accounting for the new GAIA measurements, the radial distance of 77~au becomes 64~au, with corresponding scale height of 15~au.). 

Unlike polarized differential imaging (PDI), ADI is not an ideal technique for characterizing extended disk features. As mentioned in Sect.~\ref{sec:res}, rings and spirals in protoplanetary discs appear distorted in ADI-processed images, in terms of width, intensity and position angle, especially for disc inclinations lower that 50\,\degree \citep[see][]{2012A&A...545A.111M}. 
In this work, we do not use forward modeling techniques (that can help in overcoming these distortion effects) to determine the distribution of the disc scattered light. Nevertheless, we compared our detection with previous studies in scattered light and thermal millimetre emission. 
We follow the same approach as \citet{2017ApJ...838...20M} and used a visual fit to estimate the scale height above the midplane that could reproduce our detections. 
We use the values computed in \citet{Isella2016} for the locations of the gaps and rings in the dust surface density profiles, inferred from ALMA observations. Finally, we assume a position angle and inclination of 132\degree and 42\degree, respectively. The scattering geometry is illustrated in Fig.~\ref{fig:almakeck}. 

Assuming that the arc-like feature comes from the inner edge of the first bright ring at 64~au, we find that a scattering layer at a scale height of $\sim$16~au reproduces the detection at 3.7\,$\mu$m coming from the near side of the disc in the North-East direction (see Figure \ref{fig:almakeck}, left panel). 
This estimate is obtained by shifting the ellipse corresponding to the first bright ring in the midplane along the disc minor axis, until it overlaps with the location of the extended feature in the Keck image: the offset gives the corresponding height on the midplane that is needed to explain the scattered arc. 
This value translates into an aspect ratio of $\frac{h}{r}\sim0.25$, which is more than a factor of 3 larger than found in \citet{2016A&A...588A.112G}, where the disc atmosphere was estimated to have $\frac{h}{r}\sim0.08$ outside $\sim$50~au, from modeling of sub-mm ALMA observations, corresponding to $\sim$5.1~au at a radial distance of 64~au. 
A slightly higher scale height was found by \citet{itziar} from fitting CO channel maps: the authors found $\frac{h}{r}\sim0.07$ at 1~au and flaring power of 1.12, resulting in a scale height of 7.5~au at a distance of 64~au. However, these flaring values were found to overestimate the emission at near and mid infrared wavelengths, which point to a rather flat disc \citep[see also][]{tilling2012}. 
A possible way to reconcile these discrepancies would be to assume different flaring for the inner and the outer disc. 

Applying the same reasoning to the detection at smaller separation (labeled `s' in Fig. \ref{fig:pca}), and considering the inner edge of the first gap at $\sim$36~au, we do not find a scale height that can reproduce the observed structure (see Fig. \ref{fig:almakeck}). The curvature of the detected structure deviates from the ellipse with the adopted inclination and position angle, suggesting that the emission could be due to an offset dust ring, a spiral arm, or an artefact of the image processing.

\subsection{Point source}
\label{sec:psource}
The reduced image revealed not only extended disc emission but also a point-like source (labeled `b' in Fig. \ref{fig:pca}) at $0\farcs49\pm0\farcs01$ in the North-East direction (PA=$30.6\degree\pm0.7\degree$). As explained above, we estimate a significance of 4.7$\sigma$;  
assuming Gaussian statistics we expect $\sim$10-15\% completeness for sources at this estimated flux level and angular separation.

If such emission comes from a companion, it would be located at a deprojected distance of 67~au and would therefore be close to the inner edge ($\sim70$~au) of the second gap in the dust distribution found in \citet{Isella2016}, which is a very likely site of planet formation. Assuming an age of 5$\pm$1~Myr and that the detected emission is solely the planet's intrinsic luminosity, the mass of the planet could be as high as 6.7$^{+0.6}_{-0.8}$~M$_J$ and 5.8$^{+0.7}_{-0.6}$~M$_J$ according to the Ames-COND and BT-Settl models, respectively \citep{2003A&A...402..701B,Allard2012}. On the other hand, the potential point-source may also be explained by the emission of an actively accreting circumplanetary disc with an inner radius of 1~R$_J$ and $M\dot{M}\sim 6.0\times10^{-7}$~M$_J^2$/yr \citep[based on the models computed by][]{Zhu2015}. For example, if the detected emission originates from a 0.46~M$_J$ protoplanet, which may be responsible for clearing the second gap \citep{2018ApJ...857...87L}, it is currently accreting at a rate of $\dot{M}\sim1.3\times10^{-6}$~M$_J$/yr.

\begin{figure}
\centering
\includegraphics[width=0.48\textwidth]{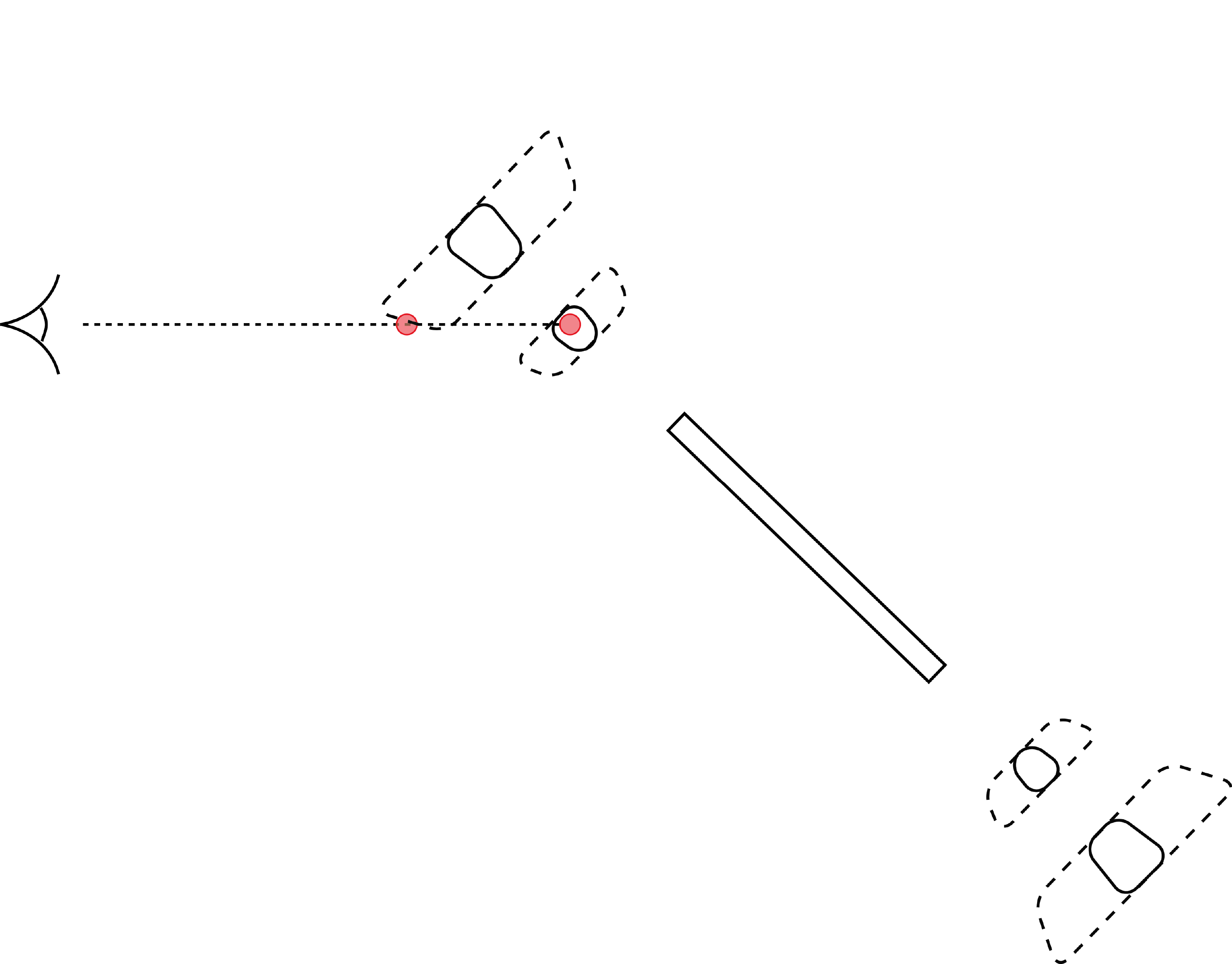}
\caption{Viewing geometry: the cartoon shows the section of HD~163296 disc inclined by 42\,\degree\hspace{1pt} with respect to the observer, which is placed on the left of the image. The first and second ring are drawn with a solid line assuming a ''flat'' disc with constant h/r=0.08; the dashed contours represent instead a scale height of 16~au above the inner edge of the first ring, and 40~au above the second ring. The blob detected at a projected distance of 0\farcs49, displayed as a red dot, could either come from a higher layer in the disc atmosphere, in correspondence to the the second bright dust-ring, or at the location of the midplane and close to the first bright ring in the $\lambda$=1.3~mm continuum.}
\label{fig:cartoon}
\end{figure}

Another possible explanation for the point-like source is scattered light from dust. 
Using the same approach as for the arc-like feature, we can estimate the height on the midplane of a scattering layer that would be necessary to reproduce the point-like source, by measuring the offset of the ellipse corresponding to the second bright ring on the midplane. 
Based on the assumed viewing geometry (see Fig. \ref{fig:cartoon}), there would need a layer $\sim$40~au above the midplane at a location of the second bright ring ($\sim$110~au) to reproduce the emission at that location in the image (Fig. \ref{fig:almakeck}), corresponding to an aspect ratio $\frac{h}{r}\sim0.36$. Considering instead the inner wall of the ring at 94~au, we find a scale height of 28~au, translating to an aspect ratio of $\frac{h}{r}\sim$0.27. 
These values are considerably higher than the scale heights found in previous studies (see Sect. \ref{sec:ext}),
and are not in good agreement with the classification of this object as a moderately flat disc \citep[either Group II, or an intermediate stage between Group~I and Group~II, see][]{2017A&A...603A..21G}. 
We note however that the unsuccessful attempts to fit  
the Spectral Energy Distribution (SED) in the near-IR and mid-IR
range using a model with continuous flaring \citep[e.g.][]{2003A&A...398..607D} and the difficulty of reproducing the observed emission of this source with a unique model from IR to millimetre \citep[e.g.][]{tilling2012,itziar} suggest a more complex structure of the disc, likely a non constant flaring and variable dust properties with radius. 

The presence of a bright spot in the arc-like feature at the same position angle (see Sect \ref{sec:ext}) as the point-like source, could point to a more efficient forward scattering in that location. Recent SPHERE observations of this target in NIR polarized light 
seem to confirm the forward scattering mechanism, with   
the first bright ring appearing brighter on the near side of the disc with respect to the far side in the PDI image \citep{2018arXiv180203328M}. The same work explained the non-detection of disc polarized scattered light outside the first bright ring with dust settling of the small grains and/or a local depletion of the smallest dust grains in the outer disc, resulting in a minimum size of 3\,$\mu$m in the grain size distribution. In this scenario we could still expect to detect the outer disc in scattered light at longer wavelengths (e.g. the L$^\prime$ band), where we are more sensitive to larger grains. 
Higher SNR observations in the NIR/mid-IR are necessary to 
determine the nature of the point-source detection, obtain further information on the grain size distribution in the disc scattering layers and rule out  possible shadowing effects of the inner disk.

\section{Conclusions}
The main goal of this program was to search for massive planets embedded in the HD~163296 disc. The Keck/NIRC2 observations revealed a clear arc-like feature that we interpret as scattered light from the disc atmosphere in the North-East region of the system, which corresponds to the near side of the disc. 
The contrast of our final frames after the PCA post-processing allowed us to set upper limits on giant planets present in this disc, at the radial position of the dust gaps found in ALMA millimetre observations. 

We also identify a point-like source at $0\farcs49\pm0\farcs01$ in the North-East side of the image at a deprojected distance of $\sim$67~au when accounting for the disc inclination and position angle. Planetary isochrones suggest that the emission may be explained by the intrinsic luminosity of a 6--7~M$_{J}$ planet \citep{2003A&A...402..701B,Allard2012}. This value is a factor of $\sim$15 larger then the planetary masses of 0.46~M$_J$ resulting from hydrodynamical simulations as presented in \citet{2018ApJ...857...87L}. However, the luminosity may not come  entirely from the planet photosphere, and the planet's mass could be much smaller if it is surrounded by an actively accreting circumplanetary disc. 
The point-like source could be also be interpreted as scattered light from dust: a forward scattering mechanism could enhance the signal coming from the near side of the disc, where the point-source is detected. 
Follow up observations in the near/mid-IR are needed to investigate the nature of the point-like source: whether it is due to a 
companion, and determine if a significant flux contribution is given by a circumplanetary disc, or on the contrary it is  scattered light from dust in the outer disc. 
Also, a more precise characterization of the disc structure may be obtained through higher resolution ALMA observations in the millimetre range, that will allow us to resolve the dust gaps and rings in order to better constrain their location on the midplane, and possibly image an accreting circumplanetary disc \citep{2018MNRAS.473.3573S}. 

\section*{Acknowledgements}
G.R. is supported by an NSF Astronomy and Astrophysics Postdoctoral Fellowship under award AST-1602444. 
The data presented herein were obtained at the W.M. Keck Observatory, which is operated as a scientific partnership among the California Institute of Technology, the University of California and the National Aeronautics and Space Administration (NASA). The Observatory was made possible by the generous financial support of the W.M. Keck Foundation. The authors wish to recognize and acknowledge the very significant cultural role and reverence that the summit of Maunakea has always had within the indigenous Hawaiian community. We are most fortunate to have the opportunity to conduct observations from this mountain. 
Part of this work was carried out at the Jet Propulsion Laboratory (JPL), California Institute of Technology, under contract with NASA. J.P.W. was supported by NASA grant NNX15AC92G. 
J.H.K. acknowledges support from NASA Exoplanets program grant NNX16AB43G to RIT. 
This research has made use of the NASA Exoplanet Archive, which is operated by the California Institute of Technology, under contract with the National Aeronautics and Space Administration under the Exoplanet Exploration Program.
Research leading to these results has received funding from the European Research Council under the European Union's Seventh Framework Programme (ERC Grant Agreement n. 337569), and from the French Community of Belgium through an ARC grant for Concerted Research Action. 
E.C. acknowledges support from NASA through Hubble Fellowship grant HF2-51355 awarded by STScI, which is operated by AURA, Inc. for NASA under contract NAS5-26555, for research carried out at the Jet Propulsion Laboratory, California Institute of Technology.
This work used archival data from HST program GO-10177 (PI: G. Schneider), which were obtained from the Mikulski Archive for Space Telescopes (MAST) at STScI, as well as data reprocessed as part of the ALICE program, which was supported by NASA through grants HST-AR-12652 (PI: R. Soummer), HST-GO-11136 (PI: D. Golimowski), HST-GO-13855 (PI: E. Choquet), HST-GO-13331 (PI: L. Pueyo), and STScI Director's Discretionary Research funds. 
A.Z. acknowledges support from the CONICYT + PAI/ Convocatoria nacional subvenci\'on a la instalaci\'on en la academia, convocatoria 2017 + Folio PAI77170087.  
This work was partly supported by the Italian Ministero dell\'\,Istruzione, Universit\`a e Ricerca through the grant Progetti Premiali 2012 -- iALMA (CUP C52I13000140001), by the Deutsche Forschungs-gemeinschaft (DFG, German Research Foundation) - Ref no. FOR 2634/1 TE 1024/1-1, and by the DFG cluster of excellence Origin and Structure of the Universe (\href{http://www.universe-cluster.de}{www.universe-cluster.de}).
\bibliography{biblio}
\bibliographystyle{mnras} 

\appendix

\section{point-like source in the post-processed frames}
\label{app:b}
\begin{figure*}
\centering
\includegraphics[width=0.47\textwidth]{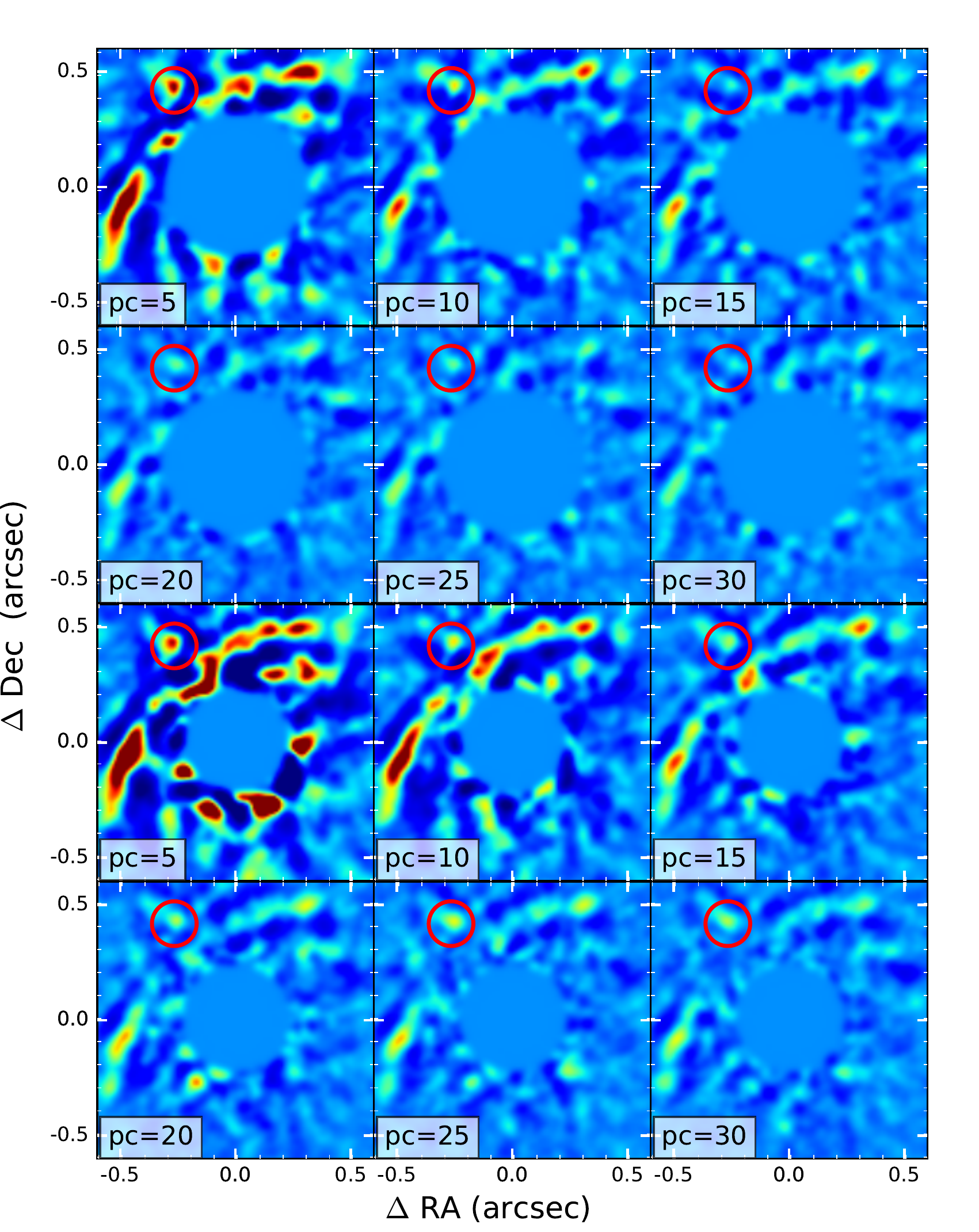}
\includegraphics[width=0.47\textwidth]{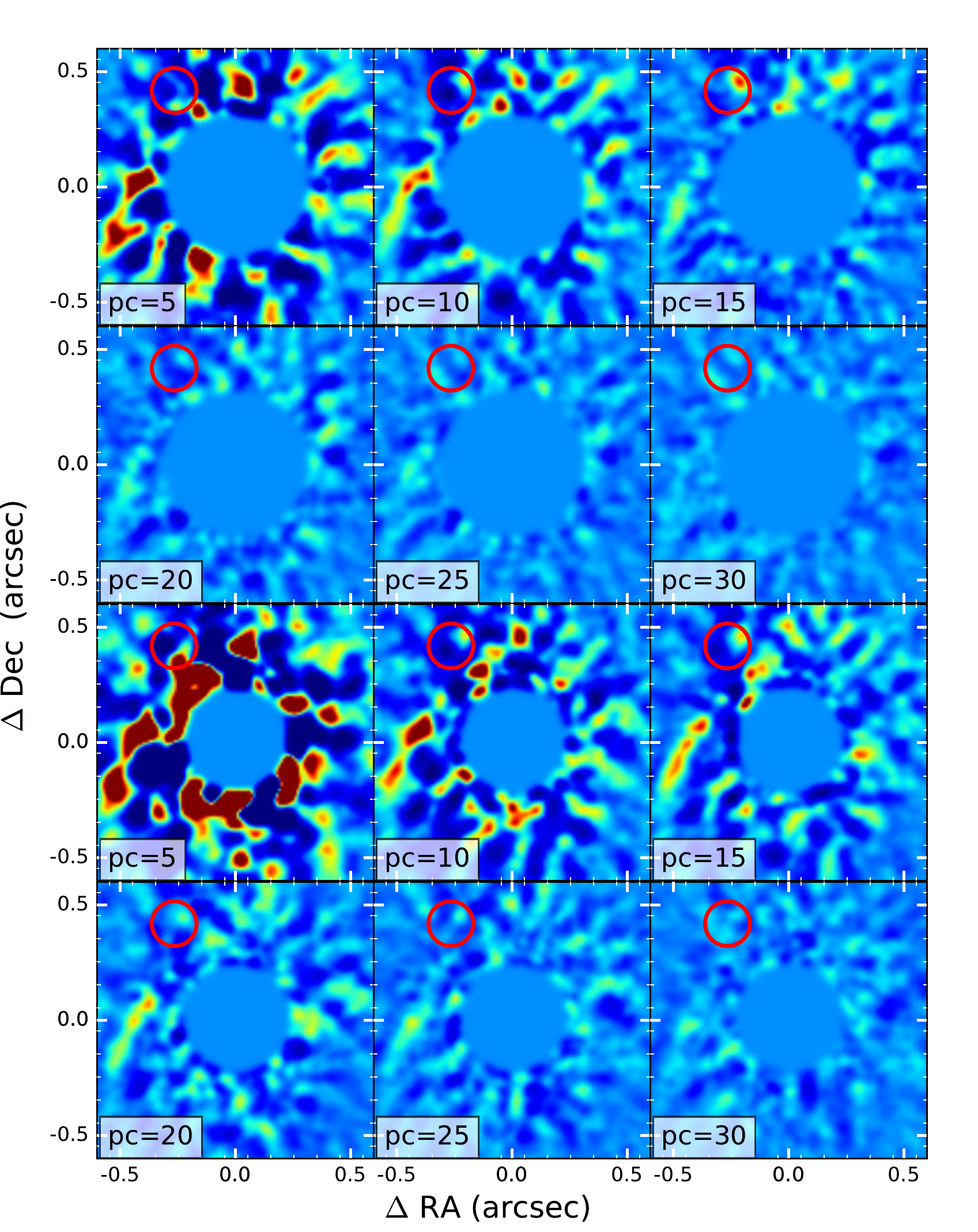}
\caption{\textit{Left panel}: PCA-ADI frames obtained from the May 30th dataset, using a inner masked region of 0\farcs32 (first and second row) and 0\farcs24 (third and fourth row), corresponding to 3 and 4 times the FWHM, respectively. The red circles are drawn at the position of the point-like source as reported in Section \ref{sec:res}. The number of principal components is noted in the box at the bottom-left of each frame. \textit{Right panel}: The same subset of PCA-ADI frames obtained with the May 31st dataset. The point-like source detection on May 30 is robust to a large range of post-processing parameters. However, we did not detect the point source on May 31, which is consistent with the relatively poor contrast limits achieved on May 31 alone (see Fig. \ref{fig:app_ccurves}). }
\label{fig:pcaframes}
\end{figure*}

\begin{figure*}
\centering
\includegraphics[width=0.55\textwidth]{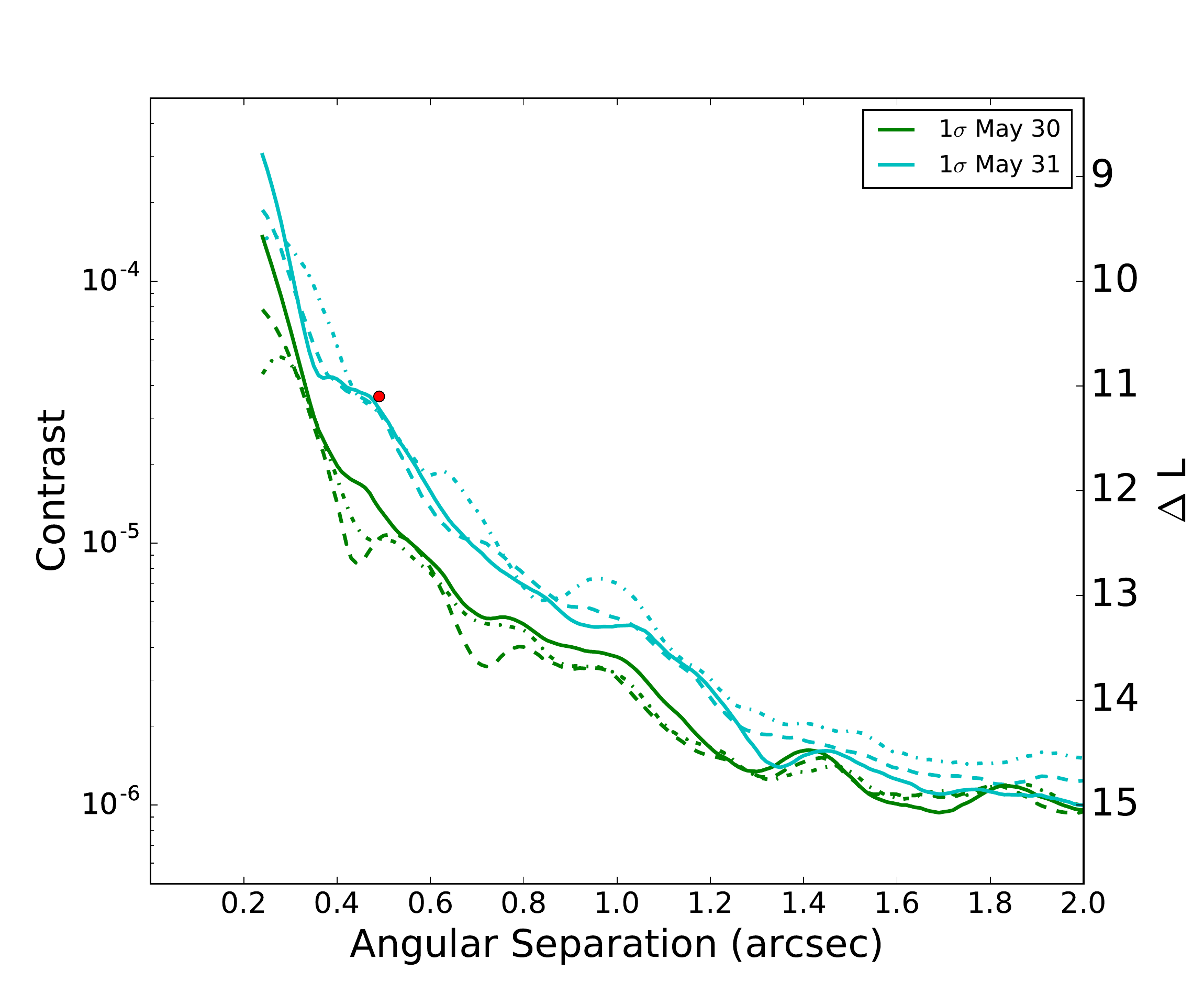}
\caption{Contrast curves at a 1~$\sigma$ level relative to a selection of frames displayed in Figure \ref{fig:pcaframes}, and computed over a parallactic angle range 140\degree--310\degree\hspace{1pt} to avoid confusion with the scattered light from the disk. The solid lines correspond to PCA-ADI with 10 principal components, the dashed lines to 20 PCs and the dashed-dotted lines to 30 PCs, and all with a inner masked region of 3*FWHM (0\farcs24). The red dot is drawn at the contrast corresponding to the point-like source described in Section \ref{sec:psource}.}
\label{fig:app_ccurves}
\end{figure*}

We show in Figure \ref{fig:pcaframes} a subset of post-processed images obtained running a full frame PCA-ADI algorithm on the separate datasets acquired during the two observing nights. The frames differ for the size of the inner masked region and the number of principal components. 
The point-like source described in section \ref{sec:psource} is consistently present in the frames obtained during the first observing night (May 30th) for a wide range of number of principal components, while the same emission is not retrieved from the second night of observation only. This is consistent with the lower quality of 
the second dataset, 
as it is noticeable from the comparison of the contrast curves, showing that the sensitivity of the frames from the May 30th observing night is systematically better than those from the second night (Figure \ref{fig:app_ccurves}). 
Similarly, the signature of the disk that is clearly visible in the frames from first Keck observing night (Fig. \ref{fig:pcaframes}, left panel) and in the GPI and SPHERE observations \citep{2017ApJ...838...20M,2018arXiv180203328M} is not so evident in the May 31st dataset alone.

\section{Background object}
\label{app:a}
We detected an additional point source at an angular separation of 2\farcs48 and PA=217$\degree$ with $\Delta L^\prime$=11.2. However, common proper motion was ruled out using archival HST/NICMOS images taken 2004-08-17, and reprocessed as part of the ALICE program \citep{Choquet2014,2018arXiv180207754H}, in which the angular separation was 2\farcs88$\pm$0\farcs04 and PA=211.0\degree$\pm0.8\degree$.

Additional sources appeared in the raw frames outside of the region used in our data reduction, with angular separations $>3\farcs75$. 

\bsp	
\label{lastpage}

\end{document}